# Spreadsheet development methodologies using Resolver. Moving spreadsheets into the 21st Century.

Patrick Kemmis
Giles Thomas
Resolver Systems Ltd
17a Clerkenwell Road, London, EC1M 5RD, UK
info@resolversystems.com

**ABSTRACT**

*We intend to demonstrate the innate problems with existing spreadsheet products and to show how to tackle these issues using a new type of spreadsheet program – Resolver. It addresses the issues head-on and thereby moves the 1980's "VisiCalc paradigm" on to match the advances in computer languages and user requirements. Continuous display of the spreadsheet grid and the equivalent computer program, together with the ability to interact and add code through either interface, provides a number of new methodologies for spreadsheet development.*

**1. INTRODUCTION**

Spreadsheets have many intrinsic problems stemming from the history of their development and the difficulties in changing the core user interface and coding language. The method of interaction has not significantly changed in over twenty years. The language used for their program code has not advanced with the developments in new dynamic computer languages.

Resolver Systems has developed a new type of spreadsheet product, which addresses many of the issues identified in this paper and offers a new method of interaction and development.

Developing robust spreadsheets is very hard using existing products. The rigorous testing of the computer software industry is absent from almost all business user-developed spreadsheets. The need to change as data requirements and calculations evolve invariably leads to their deterioration over time. The ever increasing size of data manipulated in spreadsheets only exacerbates the problems. Further, as spreadsheets are used more and more to manipulate data loaded from external databases or feeds, the inherent limitations of existing approaches are becoming clearer, and the need for alternative approaches more obvious.

One possible solution is for the problem spreadsheets to be passed to the IT departments for replacement or enhancement. However, the disconnect between business user-developed spreadsheets and IT developed programs has made it hard for solutions to cross this barrier. This has lead to spreadsheets continuing to be used well beyond their being "fit for purpose", and reluctance by IT departments to take on the conversion of spreadsheet solutions where they were not involved from the beginning.

It is proposed that these problems are reduced when the underlying program code of a spreadsheet is made visible for business or IT users to interact with, and enhance.

## 2. RESOLVER

Resolver is a new generation spreadsheet product which aims to tackle the problems identified in this paper and allow for robust and reliable spreadsheets to be developed by non IT people, while always providing a transparent bridge into an IT environment. This is achieved by changing the standard spreadsheet user interface and the underlying coding structure of the spreadsheets.

As advanced users are well aware, all spreadsheets are in reality computer programs; Resolver simply exposes to the user that underlying program alongside the traditional spreadsheet "grid". Resolver converts the spreadsheet's structure and formulae into a readable, easily-understood sequential computer program which is executed every time the spreadsheet is recalculated.

The user can input data and formulae through the traditional grid view (using normal spreadsheet data and formulae) or as user code in the "coding pane" by entering additional user defined functions and code in sections interleaved around the code generated by Resolver from the formulae and data entered into the grid. Code entered in this way can also be used by formulae entered into cells on the grid. The section where the user code is added determines its interaction with the formulae entered into the grid.

The displayed program, updated in real-time as the spreadsheet is modified, allows the user to modify and extend the program in a structured manner. It contains everything required to define the worksheets, the formatting of the cells, the data and formulae in the cells and the user code directly entered into the program code.

Resolver further displays an output window, which allows the user to follow the execution of their code, for example by printing to trace execution and to display intermediate calculation results, or by examining stack traces showing the details of any error's location.

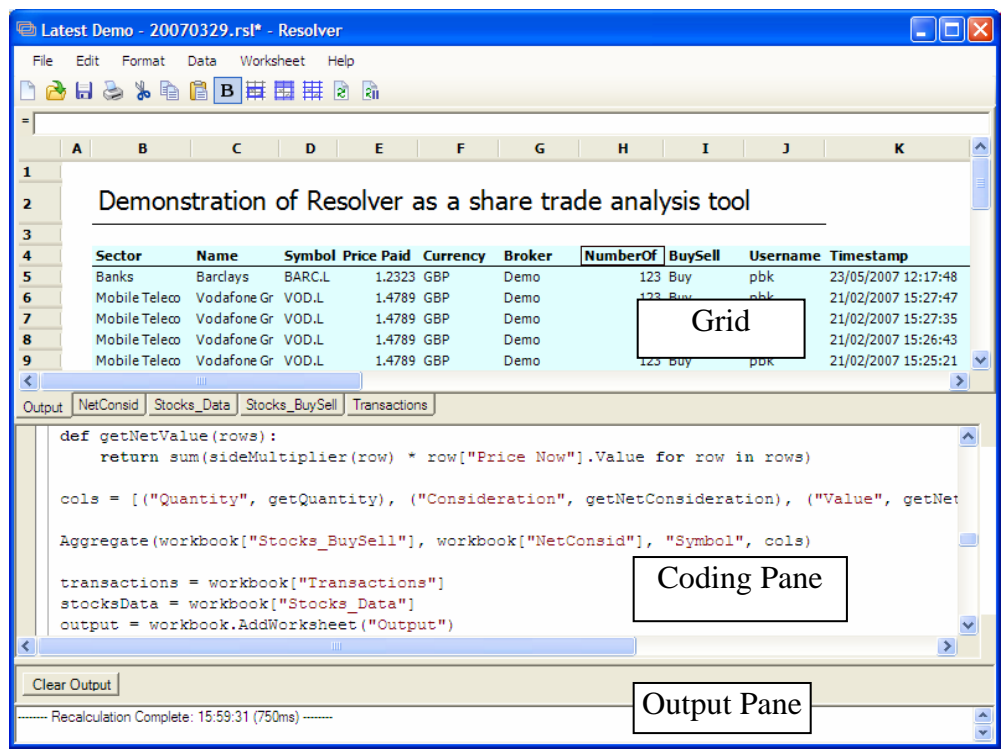

**2.1 The Coding Sections**

There are six main sections of program code which are executed by Resolver in the following order *for every recalculation* of a workbook. Three are system defined and are not editable and the other three (also thought of as *"set-up"*, *"tweak"* & *"polish"*) are user input and maintained.

**Import statements & Worksheet creation** - The code here loads appropriate background data, libraries and databases (where created through the spreadsheet user interface) and creates a workbook and worksheets for results to go into. This code is not editable by the user.

**Pre-constants user code** – The first section in the code where the user can enter their own code. It allows users to *"set-up"* their workbook as entering Python code in this section enables data from external sources to be uploaded and unique, personalised functions to be incorporated. External libraries can be loaded here from central repositories or internet sites. Bespoke links can also be established to external databases instead of using the drop-down menus to create more standard database connections. Code executed here cannot reference data or formulae entered into the grid as they have not yet been defined in the program code.

**Constants and formatting** - This section of code is generated by Resolver from cells containing just data constants (such as text or numbers but not formulae) and from all the formatting defined at cell, row or column level in each worksheet of the workbook. As data is held completely separately from the formulae, this aids locking the formulae and not the data making it easy to produce a secure data input file. This code is not editable by the user.

**Pre-formulae user code** - Entering Python code in this section enables users to *"tweak"* their calculations as it allows access to constants, but values to be defined by formulae have not yet been calculated. User code in this section can reference the constants entered into the worksheet grid, but values to be defined by formulae in the grid have not yet been calculated in each recalculation and therefore cannot be used. However functions or calculations defined in the pre-constants user code can be used.

**Formula code** - This section of code is generated by Resolver from the formulae entered into the worksheet grid. The formulae in the grid are converted where necessary to a Python compatible syntax but there is always a one-to-one relationship between formulae typed into grid cells and formulae in this section of the program code. This code is not editable by the user.

**Post-formula user code** - Entering Python code in this window can *"polish"* the result by highlighting, isolating and/or manipulating outputs for re-use elsewhere. User code in this section can use any data or results on the grid or use any functions already defined. However it cannot be used by any formulae on the grid and therefore is best used for final changes to the presentation of data on the grid or for exporting data to external files or databases etc.

**2.2 Interaction between formulae typed in the grid and user code**

Code and functions added as user code can be used by formulae entered in the grid. Similarly formulae or constants in the grid can be referenced by user code. Additional libraries and functions can be added through the code view which can be used both within the code view itself and from the cell formulae. This tightly binds the cell formulae and the code view together into a single coherent program. Defining a function in the user code allows that function to be used in a grid cell or in other user code.

An example function to add on Value Added Tax (17.5% in the UK) is shown below. This simple function - withVAT() – takes an amount passed to it (which could be a cell reference) and returns the amount after adding VAT.

```
def withVAT(amount):
    return amount*1.175
```

This function can either be used in a cell:

> If Cell A1 = 100 then typing "=withVAT(A1)" into Cell A2 would show 117.5

Or directly in user code:

> vatTotal = withVAT(workbook["Sheet1"].A2.Value)

Resolver executes the generated program (with the user's modifications and extensions) each time the user changes it, either via the spreadsheet view or directly in the code view, and then takes the results of this execution and displays them back as values in the spreadsheet or through interactions with external databases and services.

It allows the user to export the customised program - that is, both what was generated from their work in the grid, along with any code they have written - as code that can be executed as part of a traditional computer program. There is always a one-to-one relationship between the code displayed and the code executed to populate the spreadsheet.

**2.3 Program Code Language**

Resolver uses the Python programming language [van Rossum, 1995] for the program code, which is an easily learnable and highly expressive language with a rich standard library; it has been successfully used in a wide variety of applications, in both businesses and universities. Its clean syntax allows newcomers to become productive rapidly, while its more sophisticated features and built-in test framework allow experienced developers to build powerful tools which can quickly adapt to changing requirements. It has been enhanced and extended to meet the needs of a spreadsheet environment. The clear layout and syntax is in stark contrast to the complexities required to code in standard spreadsheet scripting languages.

The formula code retains compatibility with standard spreadsheet syntax but still permits the full range of Python expressions to be used offering a rich coding environment.

The formulae in the grid are converted to standard Python syntax in the program code.

Spreadsheet development using Resolver. Moving spreadsheets into the 21st Century.
Kemmis & Thomas

**2.4 The interaction of formulae & data in the grid with user & other program code**

In the very simple example below, numbers have been typed into two cells (B2, B3) and a formula into cell B4. The notes describe how the data and formula input on the grid are used in the different sections of the program code and the result is displayed back on the grid.

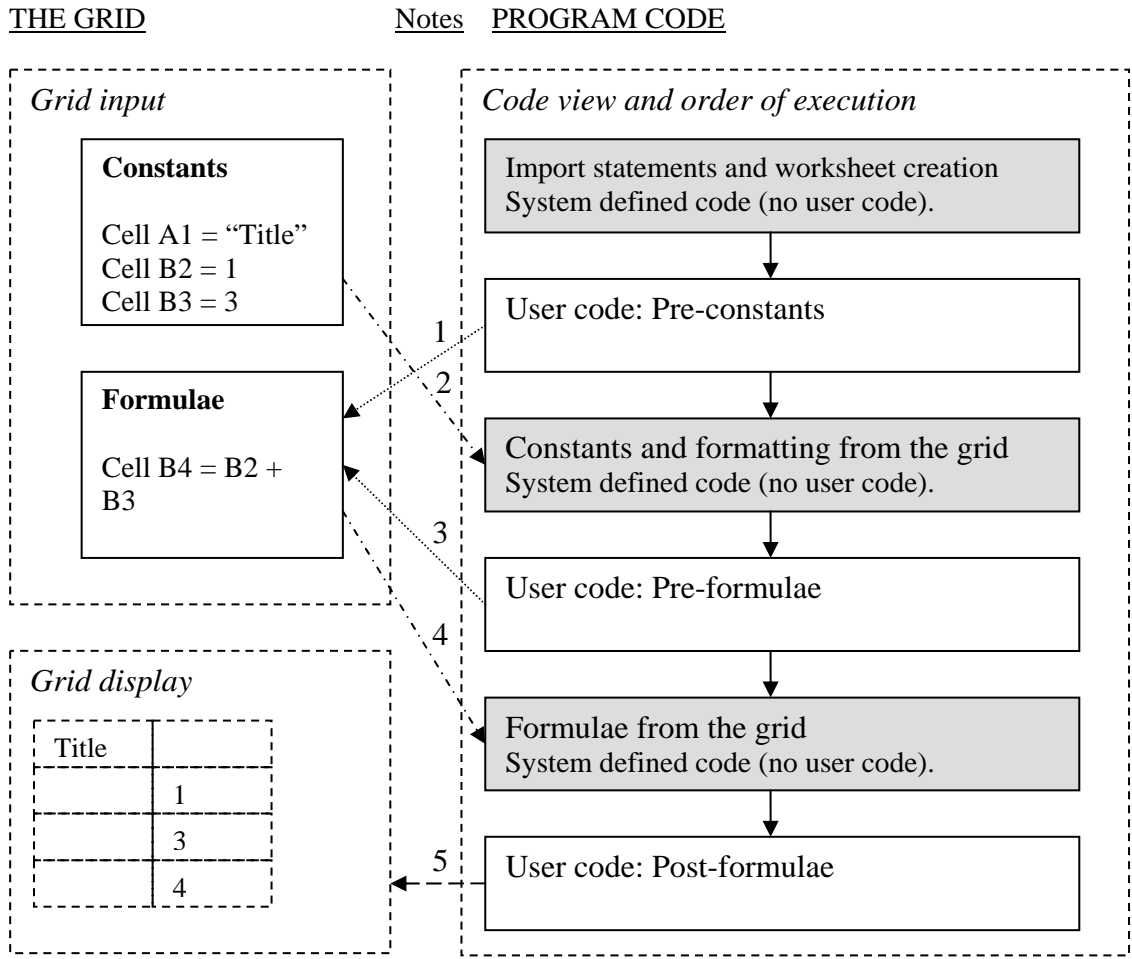

**Notes**

1. User defined functions or constants in the Pre-constants user code section are available for use by formulae entered on the grid.

2. Text and numbers entered as constants on the grid are converted automatically to program code.

3. User defined functions or constants in the Pre-formulae user code section are available for use by entered on the grid.

4. Formulae entered on the grid are converted automatically to program code.

5. The results of all stages of the program code are held in a separate copy of the grid and displayed as required.

## 3. DEVELOPMENT METHODOLOGIES

### 3.1 Loading and manipulating data from external databases

Where traditional spreadsheets are used with data supplied from external databases or other sources, the standard approach of many users is to paste the required data once into the spreadsheet and trust that it will be copied over again, should the data in the database change.

The alternative is to write a considerable amount of code to handle the data loading and later updates. However, as the shape and size of the data change over time, the code to load the data and any formulae on the worksheets all have to work reliably with the data as it changes - a not inconsiderable challenge. As data shape, volume and calculation requirements change, there is more and more opportunity for the three to get out of step - leading to errors.

Entire database tables, or the results of arbitrary SQL queries, can be imported directly into Resolver documents as new worksheets, which can immediately be used in calculations. If necessary, Resolver can update the worksheet in real time as the underlying dataset changes. Because there is no need for the user to write code themselves, this allows for sophisticated data analysis without any assistance from IT departments, raising the bar for end-user computing. Calculations can be performed based on database field references and across all rows of data ensuring that the code continues to work reliably.

### 3.2 Combining data from multiple sources

Data might logically belong in the system or model being created in the spreadsheet, but needs to be edited by more than one user simultaneously. For example, a sales director might want up-to-date information on the sales made by a number of salespeople. With a traditional spreadsheet, they might create a template for each salesperson, email a copy to each of them, and ask them to email back a completed one on a regular basis. The director would then create a summary spreadsheet with appropriate formulae to aggregate the data across all of the salespeople, and as the completed templates came back, they would copy the data over to their summary. This is clearly a very manual and error-prone system.

Resolver features Shared Worksheets, which can be simultaneously used by multiple users. Because different documents can use the same Shared Worksheet, the same data can be used for different purposes in different contexts; as the data is updated in one location, all other users can see the changes cascade through their own documents. Similarly, shared worksheets can allow a single spreadsheet to collate information from multiple sources.

### 3.3 Worksheet level formulae

Frequently worksheets are used to combine data from other worksheets; for example, the numbers in a worksheet containing account balances might be multiplied by those in equivalent positions in a worksheet containing percentage interest rates, to provide an identically-structured worksheet containing interest amounts. In a traditional spreadsheet program, specifying a worksheet like this requires a formula to be near-replicated across the entire resulting sheet, and although spreadsheets provide functionality to make this replication easier, errors can be introduced into the resulting grids of formulae easily - for example, a new row added to the end of the first worksheet and to the second worksheet will not cause a new row to be added to the end of the third.

Resolver features formulae for worksheets as well as for cells; this allows a single formula to fill a worksheet with values calculated from one or more other worksheets. Worksheets calculated in this way will be updated correctly as the input data grows or shrinks; this method also diminishes the incidence of single cells with incorrect formulae, since there is no need to mechanically specify the formula for every cell in the worksheet.

### 3.4 Working with multi-dimensional data

Standard spreadsheet products support pivot tables and other simple database methods of interacting with data where it is held against a number of dimensions. However pivot tables are notorious for being difficult to maintain and to use the resulting data.

Lotus developed a successor to Lotus 123 in the 1980s, called Improv [Zisman, 1993]; a clone called Quantrix [Rubash, A.R., Rubash M.A., 2005] is still available today. It is based on multi-dimensional data structures defined by the user, starting with a single cell, and requires the user to define all the data structures first and then populate the data afterwards. Where the requirement is to analyse pre-defined data structures such as financial Budgets and Forecasts, then this provides an easy way to "slice and dice" data across several dimensions. However, we would assert this is not how most spreadsheet users think and it has not proved to be a mainstream solution to developing general purpose robust spreadsheets. Users in general don't like having to define everything up-front, and prefer a more iterative approach to developing their spreadsheet.

Resolver allows data to be kept on separate sheets as though in a database and referenced via header rows permitting a multi-dimensional analysis to be performed as though using a database without requiring the overhead of setting up and maintaining an actual database.

### 3.5 Transfer of spreadsheet models to full programming languages

Not everything that is possible in a programming language is possible with a traditional spreadsheet; existing spreadsheet products provide separate scripting facilities that allow limited interaction with the spreadsheet model, but there is invariably a conceptual gap between the two parts of the system.

Frequently a spreadsheet used to perform a calculation must be rewritten from scratch in a traditional programming language, if it is to be reused. Business users are poor at documenting their spreadsheets and IT departments are generally unwilling to undertake the detailed analysis required to fully disassemble the code from a spreadsheet that has often evolved over time in an unstructured manner. The time required for the business user to prepare a specification, and the inevitable misunderstandings when communicating that specification from business to IT, ensures that spreadsheets continue providing strategic solutions far beyond their useful life.

Alternatively, for example, Savvysoft, a software company based in New York, has produced a product, now called Calc4Web, [Savvysoft, 2007] that converts existing spreadsheets into C++ computer code. This is great for one-off conversions of complex spreadsheet models into computer programs, but there is no seamless integration of the two, and it is not really appropriate for business user interaction with the code produced. While developing the spreadsheet model it is impossible to see the evolving computer code and the C++ code would be impenetrable to any business users anyway.

Since a Resolver document is simultaneously a spreadsheet and a computer program, the numerical data remains separate from the logic which operates on it. Resolver spreadsheets make it possible for business users to develop a spreadsheet as usual, and for IT to instantly extract their algorithms and use them in more complex systems. Equally, it allows IT departments to make live data from existing systems available to end-user spreadsheets on demand.

### 3.6 Creation of standard calculation libraries

Spreadsheet models are rarely re-used elsewhere in other spreadsheets due to the intrinsic difficulties in changing links and in documenting a spreadsheet. Often models are re-built from scratch as this is more reliable. This leads to tremendous waste and potential unintended changes. This should not be underestimated as a problem. There are many many examples of users recreating the same calculations over and over again because there is no practical method of issuing standard libraries based directly on other users' calculation models except by going through an IT development exercise [Howard, 2005].

Resolver program code derived from a spreadsheet model can be saved as a library for use in a central repository.

Resolver spreadsheets can instantly import and use any external components written for Microsoft's popular .NET platform, and many of those written for Python. This gives Resolver spreadsheets unparalleled flexibility; they can be integrated into other systems with great ease, and allow organisations to continue to make use of their existing investment in spreadsheet models.

### 3.7 Publishing to the intranet / internet

Many departments using spreadsheets need to publish the results on the internet or intranet for internal or external consumption. There is no straightforward mechanism to publish a traditional spreadsheet to a webpage.

Online spreadsheets such as Google Docs allow for easier sharing and some publishing of data, and this is great for small businesses on multiple sites or for remote access. But the user is dependent on an external online service being always available, and it does not integrate well with local spreadsheets and databases containing potentially sensitive data. This type of solution is not really appropriate for organisations requiring secure and reliable sophisticated spreadsheets where performance and control are important.

Resolver offers a web server module where any spreadsheet can be converted to a form that is immediately publishable and further provides an interaction allowing the data to be changed and re-displayed.

### 3.8 Locking down spreadsheets for data input only

There are many cases such as Budget or Forecast collection where it is imperative that the user does not change the formulae or structure of a spreadsheet so that the calculations remain correct and the central consolidation of data remains complete. This is possible with traditional spreadsheets, although the process to selectively lock cells is very cumbersome and there is no mechanism to easily extract the data and not the formulae from a worksheet.

The separation of data and formulae in Resolver permits the easy lockdown of the formulae allowing reliable use of the spreadsheet for data gathering while ensuring the ongoing reliability of the formulae. This also allows for the easy extraction of the data only for consolidation elsewhere as the constants are immediately available for consolidation.

**3.9 Data Typing**

A source of many errors in traditional spreadsheets has been through entering the wrong type of data in a cell or from a database load.

Text, dates, numbers or indeed integers all have particular needs for display and processing correctly and can have peculiar effects on formulae and therefore the results of a calculation.

There is no straightforward method to determine visually the type of data held in a set of cells. Indeed in traditional spreadsheets, all that can be determined is the display format and some properties of the data in each cell. This generally has to be determined be writing code or using the Goto function to select text or numbers.

Resolver can indicate the type of data held in every cell on a worksheet and further it can force a cell to hold and process only one type of data independently of how the data is actually displayed.

For example, numbers loaded from a database may represent codes that should be treated and sorted like text, or text that looks similar to dates needs to be processed in a single consistent manner. The ability to separate the internal processing of data from the display is key to robust and consistent treatment of data from external sources.

**3.10 Ad-hoc changes for what-if analyses**

Ad-hoc data analysis by changing values in the spreadsheet can be very useful, but sometimes changing the spreadsheet can be dangerous, as the ad-hoc changes can be accidentally saved. There is no mechanism to change the values used in a calculation while retaining the original data.

Resolver permits code created values to overwrite a user input value in a grid cell. This allows for the selective changing of data inputs for what-if analyses while still preserving the original user input data.

**3.11 Data bounds**

Because only part of a worksheet is normally visible at any one time, and there is no normal indication of whether or not there is any data "off-screen" within a worksheet, errors can be introduced by spreadsheets not taking account of data that happens to have been invisible when the user was writing part of the model. For example, if a user was summing up figures from a number of rows, a row that happened to be scrolled down below the edge of the viewable area might be missed out because there was no obvious indication that it was there. Where scripting languages are used then the problems are even worse as there is no obvious link between the size of the data and the range allowed for in the scripting code.

Resolver includes an indicator of the rectangular area bounded by the data and formulae and the coding language naturally operates over all rows or columns within the bounded area.

## 4. RESOLVER CURRENT STATUS

### 4.1 Development status

Resolver is designed to emulate standard spreadsheets as far as possible, within the new programming interface, through syntax compatibility for formulae typed in to the grid directly.  It has been in development for some eighteen months and the list of supported functions is growing all the time.

Resolver retains compatibility with traditional spreadsheets as much as possible; for example, formulae entered into the grid uses a superset of the syntax users will already be familiar with.

At the same time, the formula syntax also supports features that reflect its Python heritage - cells can contain not just numbers or text, but also lists, dictionaries, arbitrary .NET objects, and - for the truly adventurous - functions, including lambda expressions.

The code in the code view is Python pure and simple; formulae are rewritten appropriately when they are compiled down into the formula code section.

Most of the functions with which users will be familiar -such as SUM, IF, and COUNTIF - are currently supported, and the list is growing.  These have been augmented by additional functions providing a rich set of additional features such as full database access with real-time updates, access to financial data feeds such as Bloomberg, and the ability to interact with these data sets through database functionality within Resolver.

Further, the use of external publicly available libraries within the Python and the .NET environments provides further functionality.

## 5. CONCLUSION

No previous product has been really successful at combining existing spreadsheet practices with a more robust solution which can make best use of good computer software methodologies and real-time integration with the outside data world.

What is required is a product for the ordinary spreadsheet user that allows them to manage their own development of reliable spreadsheets using best practice developed in-house or externally.  It needs to allow the power user to interact with the spreadsheet using either in-cell formulae or modern simple computer code fully integrated into a whole.  The IT developer should be able to work with the same code developed by the business users without any further translation or conversion so that errors are minimised and reliability is maximised.

No previous product has combined these enhancements into one integrated solution.

Resolver as an end-user tool sits midway between traditional spreadsheets and full business intelligence tools. It still provides the familiar spreadsheet environment which business users are so familiar with, while also offering a program code view of the same data and formulae with the robust database interfaces offered by sophisticated analysis tools.

Using Resolver raises the bar of what can be developed without IT involvement, while at the same time providing an easy route for moving user-developed solutions into a full IT managed environment. It makes it possible to introduce new methodologies to spreadsheet development, where the data grid and computer language paradigms coalesce into one coherent approach.